\documentclass[a4paper,12pt]{article}
\usepackage{latexsym}

\usepackage{graphics}
\usepackage{epsf}
\usepackage{amssymb}

\catcode`\@=11
\@addtoreset{equation}{section}
\def\theequation{\thesection.\arabic{equation}}
\catcode`@=12
\relax

\newcommand{\nn}{\nonumber}

\begin{document}

%%%%%%%%%%%%%%%%%%%%%%%%%%%%%%%%%%%%%%%%%%%%%%%%%%%%%%%%%%%
%                                                         %
%  Title page                                             %
%                                                         %
%%%%%%%%%%%%%%%%%%%%%%%%%%%%%%%%%%%%%%%%%%%%%%%%%%%%%%%%%%%
\renewcommand{\thefootnote}{\fnsymbol{footnote}}
\font\csc=cmcsc10 scaled\magstep1
{\baselineskip=14pt
 \rightline{
 \vbox{\hbox{TIT-HEP-478}
       \hbox{May 2002}
       \hbox{hep-th/0205301}
}}}

\vfill
\begin{center}
{\LARGE\bf
Confining Phase Superpotentials for ${\bf SO/Sp}$ Gauge Theories via Geometric Transition
}
%\\ {\large\em --- draft \today ---}

\vfill

{\csc \large  Hiroyuki Fuji}\footnote{
      e-mail address : hfuji@th.phys.titech.ac.jp} and 
{\csc \large Yutaka Ookouchi}\footnote{
      e-mail address : ookouchi@th.phys.titech.ac.jp}\\
\vskip.1in
\vspace{0.5cm}

{\large \baselineskip=15pt
\vskip.1in
  Department of Physics, 
  Tokyo Institute of Technology,\\
  Tokyo 152-8511, Japan
}

\end{center}
\vfill

\vspace{-1cm}

\begin{abstract}
{%\baselineskip 15pt

%We confirmed the gravity dual for $\mathcal{N}=1$ $SO/Sp$ gauge
%theories 
%in terms of the confining phase superpotentials. 
%In this process, we evaluated the superpotential from the RR,NSNS flux 
%up to the 4th order of the glueball superfield. 
%And we analyzed the confining phase superpotentials 
%from purely field theory point of view using the Seiber-Witten theory. 
%Integrating out the glueball superfield from the superpotential, 
%we can compare these two superpotentials.
%
%As an example, we have considered 3 gauge groups, $SO(6),SO(8),Sp(4)$. 
%In these cases we have the perfect agreement up to the order we calculated.

We examine a large $N$ duality via geometric transition 
for ${\cal N}=1$ $SO/Sp$ gauge theories 
with superpotential for adjoint chiral superfield.
In this paper, we find that the large $N$ gauge theories are 
exactly analyzed for the classical quartic superpotentials by the finite rank 
$SO/Sp$ gauge theories.
With this classical superpotentials, 
we evaluate the confining phase superpotentials 
using the Seiberg-Witten theory.
In the dual theory, 
we calculate the superpotential generated 
by the R-R and NS-NS 3-form fluxes.
As the non-trivial examples, we discuss for $SO(6)$, $SO(8)$ and $Sp(4)$
gauge theories.
In these cases we have the perfect agreement of the confining phase 
superpotentials up to the 4th order of the glueball superfields.

}
\end{abstract}
\vfill

%\vskip.3in
\setcounter{footnote}{0}
\renewcommand{\thefootnote}{\arabic{footnote}}
\newpage

%%%%%%%%%%%%%%%%%%%%%%%%%%%%%%%%%%%%%%%%%%%%%%%%%%%%%%%%%%%%%%%%%%%%%%
\section{Introduction}
%%%%%%%%%%%%%%%%%%%%%%%%%%%%%%%%%%%%%%%%%%%%%%%%%%%%%%%%%%%%%%%%%%%%%%

It has been conjectured that 
the large $N$ gauge theory describes the string theory \cite{tHooft}.
The most prominent example of this large $N$ duality is 
$AdS/CFT$ correspondence \cite{AdS}.
As an example,
Type IIB string theory on $AdS_5\times T^{1,1}$
realizes $\mathcal{N}=1$ supersymmetric gauge theory with conformal symmetry 
in large $N$ limit, 
and the field theory results are reproduced from the supergravity 
\cite{KW}. 
In \cite{KS}, 
the breaking of the conformal symmetry of this model 
can be discussed by introducing the fractional branes.

%In the string theory, 
%the large $N$ duality
%is interpreted as open/closed string duality, and
As a large $N$ duality for topological string theory,
{\it gauge theory/geometry} correspondence is proposed 
\cite{GV0}.
Via conifold transition,
this duality claims that the topological string theory on 
the resolved conifold is equivalent to
the topological string theory on the deformed conifold
with $N$ A-branes wrapped on the special Lagrangian $3$-cycle
in the large $N$ limit.
%By introducing the other A-branes intersecting 
%with the special Lagrangian $3$-cycle, 
%the duality conjecture enables us to count the open string instanton
%\cite{OV}\cite{AV-D}\cite{OPEN}.
%Using this conjecture,
%the higher genus topological string amplitudes on
%${\cal O}(-1)\oplus{\cal O}(-1)$ bundle over ${\bf P}^1$ can be 
%evaluated from the three dimensional Chern-Simons gauge theory
%on $\mathbf{S}^3$ \cite{GV2}.
%Furthermore, the topological open string amplitudes are also 
%evaluated from the three dimensional knot invariants \cite{OV}
%and coincide with the mirror symmetry results \cite{AV}.

%This topological string duality is extended to the large $N$ duality 
%for the superstring theory \cite{Vafa} and M-theory 
%\cite{AMV}\cite{AW}.
%For Type IIA string theory, the gauge theory on $N$ D$6$-branes wrapped on 
%the $3$-cycle $S^3$ in the deformed conifold is dual to 
%the string theory on the resolved conifold with $2$-, $4$- and $6$-form 
%flux through the even homology cycles.
%As a result of this duality, 
%the superpotential for the ${\cal N}=1$ 
%$U(N)$ gauge theory coincides with that of dual gravity theory 
%in the large $N$ limit.

This topological string duality is extended to the large $N$ duality 
for the superstring theory \cite{Vafa} and M-theory 
\cite{AMV}\cite{AW}.
For Type IIB string theory, T-duality reverses the direction of the transition
\cite{Vafa}\cite{AKV}.
The duality claims that the string theory on the resolved conifold 
with $N$ D$5$-branes wrapped on the exceptional ${\bf P}^1$ is 
equivalent to that on the deformed conifold 
with $N$ units of $3$-form fluxes through the special Lagrangian 
$3$-cycle $S^3$ in large $N$ limit.
%Thus the gauge theory which is realized on D$5$-branes is 
%dual to the gravity theory with $3$-form fluxes.
%In the dual gravity theory, 
Here the fluxes on the defomed conifold
generate the superpotential and 
break the supersymmetry spontaneously \cite{TV}.
Therefore, an ${\cal N}=2$ vector multiplet splits into 
an ${\cal N}=1$ chiral and a vector multiplet.
Especially, this chiral superfield is identified with 
the glueball superfield in the confining phase of the gauge theory 
realized on the $N$ D$5$-branes in the large $N$ limit.
Under this identification, 
the superpotential generated by $3$-form fluxes
on the deformed conifold coincides with 
that of the massive glueball superfield \cite{Yank}
in the gauge theory.
In this way, the validity of the duality is examined for this geometry.

Applying the conifold transition locally, the duality is also 
considered for more complicated geometries
\cite{CIV}\cite{CFIKV}\cite{Oh-Tatar}.
In \cite{CIV}, the geometric transition for ${\cal O}(-2)\oplus{\cal O}(0)$
bundle over ${\bf P}^1$ is discussed.
In the resolved geometry, classical superpotential for $U(N)$
adjoint chiral superpotential arises in the gauge theory on $N$ D5-branes
wrapping on the exceptional ${\bf P}^1$'s, and it leads to ${\cal N}=1$ 
supersymmetric theory.
After the geometric transition, $n$ exceptional ${\bf P}^1$'s shrink 
and $n$ $S^3$'s are replaced in the dual theory.
Thus the dual theory is defined on the deformed geometry.
On this dual geometry, $3$-form fluxes appear on $S^3$'s 
after the transition 
and generate superpotential.
When the above duality conjecture is applied, the superpotential 
in the dual theory is also identified with the effective one 
for the original gauge theory in large $N$ limit.
For some examples,
the confining phase superpotentials for both theories 
coincides perfectly \cite{CIV}.

%To examine the duality for this case, 
%one needs to check the coincidence of the confining phase superpotentials 
%on both sides.
%For $U(N)$ gauge group, 
%the large $N$ gauge theory with the classical cubic superpotential
%can be exactly analyzed.
%In the dual geometry, the effective superpotential is evaluated from 
%the periods.
%As a result, these superpotentials coincide.

%In \cite{CFIKV},
%the duality is extended to 
%$\mathcal{N}=1$ quiver theories on D5 branes 
%wrapped on the exceptional ${\bf P}^1$'s of threefold 
%which is ALE fibration over a complex plane.
%On these geometries,
%the dual gravity theories naturally realize 
%the duality cascades in ${\cal N}=1$ gauge theory.

%One can also prove
%the coincidence of the confining phase superpotentials in this case.
%In \cite{CKV}\cite{CFIKV}, 
%$\mathcal{N}=1$ quiver theories on 
%D5 branes wrapped on the exceptional 2-cycles of threefolds 
%which are ALE fibrations over a complex plane was constructed
%and 
%then discussed geometric transitions for these models. 
%The fibration of ALE space is classified into 
%the monodormic fibration and the nonmonodormic fibration. 
%In particular the model with two adjoint fields have been constructed 
%from a monodormic fibration. 
%Seiberg-like dualities have been viewed in a unified way \cite{CFIKV}. 
%Also duality cascades are naturally realized in these classes
%of theories, and are related to the affine Weyl group symmetry. 

By introducing orientifold projection, the conifold is resolved by
${\bf RP}^2$ and the gauge group becomes $SO(N)/Sp(N)$.
The large $N$ duality of unoriented string is examined 
for the topological string theory \cite{SV}\cite{AAHV}
and Type IIB superstring theory \cite{EOT}\cite{IY}\cite{Gomis}.
In  \cite{EOT},
the geometric transition for $SO(N)$ gauge theory is discussed 
in this set up.
As in $U(N)$ gauge theory case,
The world-volume theory of $N$ D5-branes wrapped on ${\bf RP}^2$  
realize $SO(N)/Sp(N)$ gauge theory with arbitrary classical superpotential 
for the adjoint chiral superfield.
In order to confirm the duality proposal for this case,
it is also necessary to examine the coincidence of the physical
quantities on both theories.
In \cite{EOT}, it is found that 
the effective coupling constant of $SO(N)$ gauge theory 
agrees with that of the dual theory.

%To examine the duality for $SO/Sp$ gauge theories further,
%the other physical quantities must also be coincident with that of the dual 
%theories.
In order examine the duality for $SO/Sp$ gauge theories further,
we evaluate the confining phase superpotentials 
on both theories in this paper.
On the gauge theory side,
we need to evaluate it in large $N$ limit of $SO(N)/Sp(N)$ gauge group.
We prove that 
it can be evaluated exactly from the {\rm finite} rank gauge theory 
for the $SO/Sp$ gauge theory
with the classical quartic superpotential for adjoint chiral superfield. 
On the dual theory side, 
we evaluate the periods for the deformed geometry with 
the orientifolding.
By identifying the periods with the expectation values 
of the gluon superfields, we obtain the effective superpotentials.
To evaluate the superpotentials explicitly on both sides,
we consider $SO(6)$, $SO(8)$ and $Sp(4)$
gauge theories as the non-trivial examples.
As a result, we find the perfect agreement of these superpotentials 
up to the 4th order of the glueball superfields.

%This method was discussed in \cite{CIV} for $U(N)$ case. 
%We confirmed the gravity dual for $SO/Sp$ gauge theories 
%in terms of the method. 
%In this process, we evaluated the superpotential form the RR,NS flux 
%up to the 4th order of the glueball superfield. 
%And we analyzed the confining phase superpotentials 
%form purely field theory point of view using the Seiberg-Witten theory. 
%Integrating out the glueball superfield form the superpotential, 
%we can compare these two superpotentials.
%As an example, we have considered 3 gauge groups, $SO(6),SO(8),Sp(4)$. 
%In these cases we have the perfect agreement up to the order we calculated.

This paper is organized as follows.
In section 2, we will review the geometric transition \cite{Vafa}
and see how the gauge theory and the string theory are exactly 
analyzed \cite{CIV}\cite{EOT}.
In section 3, 
we will prove that the confining phase superpotential for $SO(2N)/Sp(2N)$ 
gauge theory in the large $N$ limit is 
exactly evaluated from the finite rank gauge theory with the classical 
quartic superpotential.
And then we will compute the confining phase superpotential 
using the Seiberg-Witten theory for the gauge groups $SO(6)$, $SO(8)$
and $Sp(4)$.
In section 4, we will explicitly analyze the dual geometry and evaluate 
the effective superpotential from the computation of the periods. 
By comparing the results of section 3 and section 4, 
we will find the coincidence of the exact superpotentials.
In the Appendix we will show the detailed computations of periods in section 3.

%%%%%%%%%%%%%%%%%%%%%%%%%%%%%%%%%%%%%%%%%%%%%%%%%%%%%%%%%%%%%%%%%%%%%%
\section{Geometric Transition and Large $N$ Duality}
%%%%%%%%%%%%%%%%%%%%%%%%%%%%%%%%%%%%%%%%%%%%%%%%%%%%%%%%%%%%%%%%%%%%%%
\subsection{The Geometric Transition}

%%%%%%%%%%%%%%%%%%%%%%%%%%%%%%%%%%%%%
%Resolved geometry and N=1 SYM
%%%%%%%%%%%%%%%%%%%%%%%%%%%%%%%%%%%%%

%{\bf geometric engineering}
 
We consider Type IIB string theory on non-compact Calabi-Yau manifold,
%$4$ dimensional ${\cal N}=1$ supersymmetric $U(N)$ gauge theory 
%can be realized on $N$ D$5$-branes
%which wraps on the exceptional ${\bf P}^1$ 
%in the non-compact Calabi-Yau manifold,
\begin{eqnarray}
M_s:\,W^{\prime}(x)^2+y^2+z^2+w^2=0,
\label{ALE}
\end{eqnarray}
where $W(x)$ is defined as,
\begin{eqnarray}
W(x)\equiv \sum_{p=1}^{n+1}\frac{g_p}{p}x^{p}.
\label{W(x)} 
\end{eqnarray}
This Calabi-Yau manifold is resolved by 
locating ${\bf P}^1$'s at the singularities where
$W^{\prime}(x)=0$ is satisfied,  
and defined as 
${\cal O}(-2)\oplus {\cal O}(0)$ bundle over ${\bf P}^1$.
When $N$ D$5$-branes
wraps on the exceptional ${\bf P}^1$'s and fill the flat $4$ dimensional 
space-time,
$4$ dimensional ${\cal N}=1$ supersymmetric $U(N)$ gauge theory 
with superpotential $W_{{\rm cl}}$ for the adjoint chiral superfield $\Phi$, 
\begin{equation}
W_{\rm cl}(\Phi)=\sum_{p=1}^{n+1}\frac{g_p}{p}{\rm Tr}\Phi^p,
\end{equation}
is realized \cite{Katz}.

In order to realize $SO(N)/Sp(N)$ gauge theories, we introduce orientifold projection which is complex conjugation. For singular geometry (\ref{ALE}) this projection acts as
\begin{equation}
(x,y,z,w) \to (\bar{x},\bar{y},\bar{z},\bar{w}).
\label{complex conjugation}
\end{equation}
To extend this projection to the resolved geometry we consider new variables,\begin{eqnarray}
a=x+iy,\ \  b=z+iw,\ \  c=x-iy,\ \  d=-z+iw.
\end{eqnarray}
The complex conjugation (\ref{complex conjugation}) corresponds to the following action,
\begin{eqnarray}
a\to \bar{c},\ \ b\to -\bar{d},\ \ c\to \bar{a},\ \ d\to -\bar{b}.
\label{newconju}
\end{eqnarray}
The resolved conifold can be described as a union of two patch, $M_1=\{(a,b,c,z)|d=ac\}$ and $M_2=\{(a^{\prime},b^{\prime},c^{\prime},z^{\prime})|c^{\prime}=b^{\prime}d^{\prime}\}$ glued together by the identification $b=a^{\prime}b^{\prime}$. The singular point of the Calabi-Yau manifold is replaced by ${\bf P^1}$ whose coordinates are given by $a^{\prime}$ on $M_1$ and $b^{\prime}$ on $M_2$. By the orientifolding (\ref{newconju}) we identify $a^{\prime}$ with $-\bar{b}^{\prime}$. This is antipodal map, then ${\bf P^1}$ becomes ${\bf RP^2}$.

In this geometry world-volume theory on the D5-brane is $SO/Sp$ gauge theory with the following tree level superpotential \cite{EOT},
\begin{equation}
W_{\mathrm{cl}}(\Phi)= \sum_{p=1}^{n+1}\frac{g_{2p}}{2p}{\rm Tr}\Phi^{2p}
\equiv\sum_{p=1}^{n+1}g_{2p}u_{2p}, \label{classical pote}
\end{equation}
where $\Phi$ is the chiral superfield in the adjoint representation 
of $SO(N)/Sp(N)$ gauge group and $u_{2p}\equiv\frac{1}{2p}{\rm Tr}\Phi^{2p}$.

We define parameters $a_i$ by
\begin{eqnarray}
W^{\prime}(x)=\sum_{p=1}^{n}g_{2p}x^{2p-1}
=g_{2n+2}\ x\prod_{i=1}^{n}(x^2+a_i^2). \label{classical sup}
\end{eqnarray}
%An exceptional ${\bf P}^1$ on which a  orientifold plane 
%wraps is replaced by ${\bf RP}^2$ at $x=0$. 
%The other ${\bf P}^1$ located at $x=ia_p$ is mapped to 
%${\bf P}^1$ located at $x=-ia_p$. 
%Thus the numbers of D$5$-branes which wrap these pair of ${\bf P}^1$'s
%must be same.

%On flat 4 dimensional world volume of $N$ D5-branes 
%and a O$5$-plane, 

In the classical vacua of this gauge theory, 
the eigenvalues of $\Phi$ become roots $0,\pm ia_i$'s of $W^{\prime}(x)=0$.
When $N_0$ D5-branes wrap on ${\bf RP}^2$ and 
$N_i$ D5-branes wrap on the ${\bf P}^1$ located at $x=\pm ia_i$, 
the vacuum of the gauge theory becomes classically 
$P(x)\equiv{\rm det}(x-\Phi)=x^{N_0}\prod_{i=1}^{n}(x^2+a_i^2)^{N_i}$
and the gauge group breaks as,
\begin{eqnarray}
SO(N)\to SO(N_0)\otimes_{i=1}^{n}U(N_i),\quad
Sp(N)\to Sp(N_0)\otimes_{i=1}^{n}U(N_i),
\nn
\end{eqnarray}
where $N=N_0+\sum_{i=1}^{n}N_i$.
This theory can be analyzed non-perturbatively 
in terms of the Seiberg-Witten theory \cite{EFGIR}\cite{TY} 
and we will review in the next section.

%%%%%%%%%%%%%%%%%%%%%%%%%%%%%%%%%%%%%%%%%%%%%%%
%Geometric Transition and Deformed Geometry
%%%%%%%%%%%%%%%%%%%%%%%%%%%%%%%%%%%%%%%%%%%%%%%
\vspace{0.5cm}

%{\bf geometric transition}

The large $N$ dual of this theory is found via conifold transition
\cite{GV0}\cite{Vafa}.
The conifold is defined as \cite{Candelas}\cite{strominger},
\begin{eqnarray} 
x^2+y^2+z^2+w^2=0.
\end{eqnarray}
This geometry has a singularity at $x=y=z=w=0$.
This singularity can be removed in two ways.
One way is the deformation of K\"ahler structure. 
This singularity is 
blown up and replaced by ${\bf P}^1$.
The resulting geometry becomes ${\cal O}(-1)\oplus{\cal O}(-1)$ bundle 
over ${\bf P}^1$,
and this is called resolved conifold. 
Another way is the deformation of the complex structure, where 
the defining equation is deformed as,
\begin{equation}
x^2+y^2+z^2+w^2=\mu^2.
\label{deformed conifold}
\end{equation}
Thus the singularity is replaced by $S^3$ with the radius $\mu$
and resulting geometry becomes $T^{*}S^3$.
This manifold is called deformed conifold.
When $N$ D5-branes are wrapped on ${\bf P}^1$ in the resolved conifold, 
D5-branes disappear and $N$ units of R-R $3$-form flux $H_{R}$ 
through $S^3$ and NS-NS $3$-form flux $H_{NS}$ 
through the dual $3$-cycle remain after the conifold transition. 
In the large $N$ limit, these two theories are equivalent and 
give the same physical quantities \cite{Vafa}.
%When a O$5$-plane is introduced into the resolved conifold,
%the anti-podal identification makes
%the exceptional ${\bf P}^1$ to ${\bf RP}^2$ as discussed above.
%On the other hand, the deformed conifold (\ref{deformed conifold})
%is invariant under the complex conjugation (\ref{complex conjugation}).
%Therefore the dual geometry is also given by the deformed conifold.
After the conifold transition, D$5$-branes and orientifold plane disappear and 
$N\mp 2$ units\footnote{The sign $\mp$ is determined by the 
sign of the charge of orientifold plane.}
of R-R $3$-form flux $H_{R}$ 
through $\mathbf{S}^3$ and NS-NS $3$-form flux $H_{NS}$ 
through the dual $3$-cycle 
remain \label{SV}.

In the case of the Calabi-Yau manifold (\ref{ALE}), 
above analysis can be applied locally.
Through the geometric transition, the dual Calabi-Yau manifold  
is defined by replacing all ${\bf P}^1$'s and a ${\bf RP}^2$ by $S^3$'s. 
The deformed geometry is the following hypersurface in ${\bf C}^4$, 
\begin{equation} 
M_{{\rm cpx}}:\,g \equiv W^{\prime}(x)^2+f_{2n-2}(x)+y^2+z^2+w^2=0,
\label{deform ALE}
\end{equation}
where $f_{2n-2}(x)$ is the degree $(n-1)$-th polynomial of $x^2$.
In this deformed geometry, the integral basis of the 
$3$-cycles $A_i,B_i\in H_3(M,{\bf Z})$ ($i=1,\cdots,h_{2,1}=2n+1$) 
satisfy the symplectic pairing,
\begin{equation}
(A_i , B_j)=-(B_j ,A_i)=\delta_{ij}\ ,\qquad (A_i,A_j)=(B_i,B_j)=0,
\end{equation}
where the pairing $(A,B)$ of three-cycles $A,B$
is defined as the intersection number.
For the deformed Calabi-Yau manifold (\ref{deform ALE}),
these $3$-cycles are constructed as $\mathbf{P}^1$ fibration over the 
line segments between two critical points 
$x=0^{+},0^{-},\pm ia_1^{+},\pm ia_1^{-}\cdots 
\pm ia_{n}^{+},\pm ia_{n}^{-},\infty$ of $W^{\prime}(x)^2+f_{2n-2}(x)$
in $x$-plane.
%Since ${\bf Z}_2$ symmetry is imposed on this geometry, 
%we will concentrate on the upper half plane of $x$-plane in the following.
Therefore we set the three cycle $A_0$ to be the ${\bf P}^1$ fibration 
over the line segment between $0^-$ and $0^{+}$ and 
three cycle $ A_i$ to be the fibration 
over the line segment between $ia_i^-$ and $ia_i^{+}$.
On the other hand, 
three cycle $B_0$ is constructed as ${\bf P}^1$ fibration 
over the line segment between $0^+$ and $\Lambda_0$ and 
three cycle $B_i$ to be the fibration 
over the line segment between $ia_i^+$ and $i\Lambda_0$, 
Here we introduced the cut-off $\Lambda_0$, 
as these cycles are non-compact.
Since this geometry has $\mathbf{Z}_2$ symmetry, the discussion is  
restricted to the upper half of $x$-plane in the following \cite{EOT}.

%%%%%%%%
 \begin{figure}[ht]
 \centerline{\epsfxsize=8cm \epsfbox{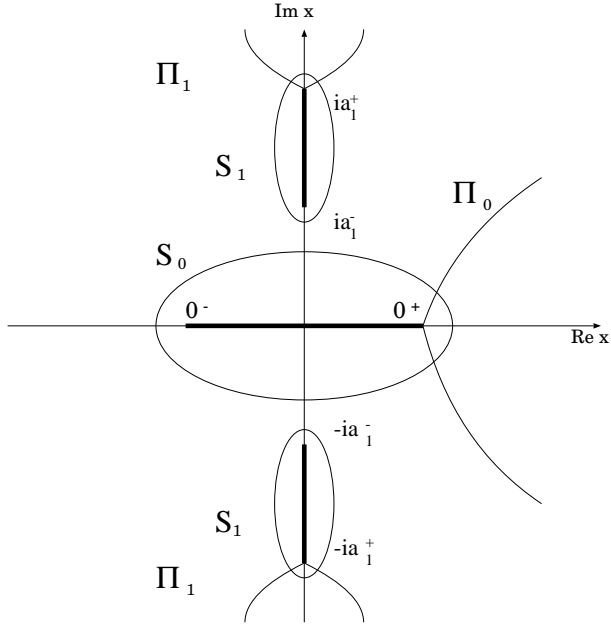}}
  \vskip 3mm
  \caption{The 3-cycles which are constructed as the ${\bf P}^1$ fibration 
           over the line segments in the complex plane.}
 \end{figure}
%%%%%%%%

The holomorphic $3$-form $\Omega$ for the deformed geometry 
(\ref{deform ALE}) is given by
\begin{equation}
\Omega=2\frac{dx\wedge dy\wedge dz}{\partial g/\partial w}.
\end{equation}
The periods $S_i$ and dual periods $\Pi_i$ for 
this deformed geometry is given as,
\begin{equation}
S_i=\int_{A_i}\Omega, \quad \Pi_i=\int_{B_i}\Omega.
\end{equation}
The dual periods are expressed in terms of the prepotential
${\mathcal F}(S_i)$ such as,
\begin{equation}
\Pi_i=\partial {\cal F}/\partial S_i.
\label{prepotential}
\end{equation}
Since these $3$-cycles are constructed as ${\bf P}^1$ fibrations, 
these periods are written in terms of the integrals over $x$-plane as,
\begin{eqnarray}
&&
S_0=\frac{1}{2\pi i}\int_{0^{-}}^{0^{+}}\omega,\quad
S_i=\frac{1}{2\pi i}\int_{ia_i^{-}}^{ia_i^{+}}\omega,
\nn \\
&&
\Pi_0=\frac{1}{2\pi i}\int_{0^{+}}^{\Lambda_0}\omega,\quad
\Pi_i=\frac{1}{2\pi i}\int_{ia_i^{+}}^{i\Lambda_0}\omega,
\end{eqnarray}
where $\omega$ is obtained by integrating holomorphic $3$-form 
over the fiber ${\bf P}^1$,
\begin{equation}
\omega=2dx\ \left(W^{\prime}(x)^2+f_{2n-2}(x)\right)^{\frac{1}{2}}.
\label{1-form}
\end{equation}

%%%%%%%%%%%%%%%%%%%%%%%%%%%%%%%%%%%%%%%%%%%%%%%%%%%%%%%%%%%%%%%%%%%%%%%
\subsection{Partial SUSY Breaking and Confinement of Gauge Theory}

When the geometric transition occurs, 
the exceptional ${\bf P}^1$'s on which 
D$5$-branes wrap in the resolved geometry, 
is replaced by $3$-form fluxes through the special Lagrangian 
$3$-cycles in the deformed geometry.
This $3$-form fluxes generate the superpotential, and 
${\mathcal N}=2$ supersymmetry for the dual theory 
is broken partially to ${\mathcal N}=1$ supersymmetry \cite{TV}.
In this subsection, we will review how the partial supersymmetry breaking 
occurs in the dual geometry
and how the supermultiplets in the dual theory are
identified with that of the effective gauge theory 
by the large $N$ duality conjecture \cite{Vafa}.

The partial supersymmetry breaking of ${\cal N}=2$ theory to 
${\cal N}=1$ theory occurs by the electric and
magnetic Fayet-Iliopoulous superpotential terms as \cite{Antoniadis},
\begin{equation}
W_{\mathrm{FI}}=\sum_{i}\int d\theta^2 d\tilde{\theta}^2\left(e_i\Psi_i
+m_i\frac{\partial{\cal F}}{\partial \Psi_i}\right),
\end{equation}
where 
$e_i$'s and $m_i$'s are electric and magnetic charge respectively, and
$\Psi_i$'s are ${\cal N}=2$ superfields and 
the holomorphic function ${\cal F}$ is the prepotential
for the ${\cal N}=2$ theory.

Turning on R-R and NS-NS fluxes through the special Lagrangian $3$-cycles,
the above partial supersymmetry breaking is realized 
in Type IIB string theory \cite{TV}.
The $3$-form fluxes generate the superpotential \cite{GVW}
\begin{equation}
-\frac{1}{2\pi i}W_{{\rm eff}}
=\int\Omega\wedge(H_{{\rm R}}+\tau H_{{\rm NS}}),
\label{flux pot}
\end{equation}
where $H_{R}$ and $H_{NS}$ are $3$-form fluxes and $\tau$ is the
complexified Type IIB string coupling, and $\Omega$ is the holomorphic 
$3$-form on the Calabi-Yau manifold.
In the case of dual theory defined through geometric transition, 
$H_{\rm R}$ and $H_{NS}$ satisfy,
\begin{eqnarray}
N_0\mp 2=\int_{A_0}H_{{\rm R}}, \quad 
N_i=\int_{A_i}H_{{\rm R}}, \quad 
\alpha=\int_{B_i}H_{{\rm NS}},
\end{eqnarray}
where $\alpha$ is the $4$ dimensional 
bare gauge coupling constant $g_0$ as 
$\alpha\equiv 4\pi i/g_0^2$.

Plugging these relations into (\ref{flux pot}), 
the superpotential for the dual theory is expressed 
in terms of periods $S_i$ and dual periods $\Pi_i$ of 
the deformed Calabi-Yau manifold such as,
\begin{equation}
-\frac{1}{2\pi i}W_{{\rm eff}}
=\left(\frac{N_0}{2}\mp1\right)\Pi_0
+\sum_{i=1}^{n}N_i\Pi_i+\alpha\sum_{i=0}^{n}S_i. 
\label{superfulx}
\end{equation}
With this superpotential, ${\cal N}=2$ vector multiplets $\Psi_i$ 
splits into the massive ${\cal N}=1$ chiral superfields $S_i$ 
and massless $U(1)^n$ vector multiplets. 
Following the large $N$ duality proposal \cite{Vafa}, 
massless $U(1)^n$ vector multiplets are identified with those
in the effective theory of the 
${\cal N}=1$ $SO(N)/Sp(N)$ gauge theory with the classical superpotential 
$W_{{\rm cl}}(\Phi)$.
The ${\cal N}=1$ massive chiral superfield $S_i$ is  
identified with the glueball superfield,
\begin{equation}
S_i=-\frac{1}{32\pi^2}{\rm Tr}_{SU(N_i)}W_{\alpha}W^{\alpha},
\end{equation}
where $W_{\alpha}$ is defined as $W_{\alpha}\equiv D_{\alpha}V$ with
supercovariant derivative $D_{\alpha}$
and ${\cal N}=1$ vector multiplet $V$.
Thus the dual theory on the deformed geometry 
with fluxes corresponds to 
the confining phase of the gauge theory which is determined in terms of
the resolved geometry.

%To examine this correspondence, we must see the properties of the 
%gauge couplings.
%The effective $U(1)^{n}$ gauge coupling $\tau_{ij}$ is given by
%\begin{equation}
%\tau_{ij}=\partial \Pi_i/\partial S_j
%=\partial^2 {\cal F}/\partial S_i \partial S_j.
%\end{equation} 
%and this is consistent with the gauge theory
%in the confining phase \cite{SW}.
%Therefore the gauge couplings for the $U(1)^n$ gauge theory 
%determined in terms of the deformed geometry can be identified 
%with the gauge couplings for the effective theory of the 
%$SO(N)/Sp(N)$ gauge theory determined in terms of the resolved geometry. 

To examine this correspondence, we should check that 
the low energy superpotential $W_{\rm exact}$
in the confining phase of ${\cal N}=1$ $SO(N)/Sp(N)$ gauge theory 
%with the superpotential $W_{{\rm cl}}(\Phi)$
%which is constructed from the resolved geometry with $N$ D5-branes, 
coincides with the superpotential $W_{{\rm eff}}(S_i)$ which is generated by 
the deformed geometry with $3$-form fluxes.
In the following sections, we will evaluate these superpotentials and 
see their coincidences.

%%%%%%%%%%%%%%%%%%%%%%%%%%%%%%%%%%%%%%%%%%%%%%%%%%%%%%%%%%%%%%%%%%%%%%
\section{Confining Phase Superpotentials}
%%%%%%%%%%%%%%%%%%%%%%%%%%%%%%%%%%%%%%%%%%%%%%%%%%%%%%%%%%%%%%%%%%%%%%
\subsection{Confining Phase Superpotentials for $SO/Sp$ SYM}

The confining phase of the pure ${\cal N}=2$ gauge theory 
is analyzed by the Seiberg-Witten theory \cite{SW}.
For the gauge group $SO(2N)$, the Seiberg-Witten curve is written as,
\begin{equation}
y^2=\left[P_{2N}(x^2,u_i)\right]^2-4x^4\Lambda^{4N-4},
\end{equation}
where the characteristic polynomial $P_{2N}$ is defined as \cite{9505088},
\begin{eqnarray}
&&P_{2N}(x^2,u_i)\equiv{\rm det}(x-\Phi)=\sum_{k=0}^{2N}x^{2N-2k}s_{2k},
\nn \\
&&\Phi\equiv{\rm diag}\left(
ia_1\sigma_2,ia_2\sigma_2,\cdots,ia_N\sigma_2
\right).
\end{eqnarray}
Here $\sigma_a$, $(a=1,\cdots,3)$ are the Pauli matrices 
and $s_{2k}$'s satisfy the 
following Newton's relations,
\begin{eqnarray}
ks_{2k}+\sum_{r=1}^{k}ru_{2k}s_{2k-2r}=0,
\quad
u_k\equiv \frac{1}{k}{\rm Tr}\Phi^k.
\label{Newton}
\end{eqnarray}
For the gauge group $Sp(2N)$, the Seiberg-Witten curve 
is written as \cite{9509175}\cite{9504102},
\begin{equation}
x^2y^2=\left[P_{2N}(x^2,u_i)+2\Lambda^{2N+2}\right]^2
-4\Lambda^{4N+4},
\end{equation}
where the characteristic polynomial $P_{2N}$ is defined for the 
adjoint superfield $\Phi$ which is defined as,
\begin{eqnarray}
&&
^t\Phi=J\Phi J, \quad
J={\rm diag}(i\sigma_2,\cdots,i\sigma_2),
\nn \\
&&
J\Phi={\rm diag}(a_1\sigma_1,\cdots,a_N\sigma_1).
\end{eqnarray}

When the superpotential $W_{{\rm cl}}(\Phi)$ is introduced, 
the ${\cal N}=2$ gauge theory is deformed and the resulting theory 
has the unbroken supersymmetry on the submanifold of the Coulomb branch.
This submanifold is determined by the locus where Seiberg-Witten curve 
degenerates. 
The monopoles or dyons become massless on some particular 
submanifold $\langle u_{2k}\rangle$.
Near a point with $l$ massless monopoles, the superpotential is,
\begin{eqnarray}
W=\sum_{k=1}^{l}M_{k}(u_{2r})q_{k}\tilde{q}_{k}
+\sum_{p=1}^{n+1}g_{2p}u_{2p}.
\end{eqnarray}
On the supersymmetric vacua, $\langle u_{2k}\rangle$'s satisfy,
\begin{equation}
M_{k}(\langle u_{2k}\rangle)=0,
\quad
g_{2k}+\sum_{p=1}^{l}
\frac{\partial M_p(\langle u_{2k}\rangle)}{\partial \langle u_{2k}\rangle}
\langle q_{p}\tilde{q}_{p}\rangle=0.
\end{equation}
Therefore the superpotential in this vacuum is simply \cite{EFGIR}\cite{TY},
\begin{equation}
W_{{\rm exact}}=\sum_{p=1}^{n+1}g_{2p}\langle u_{2p}\rangle.
\end{equation}

In the confining phase where 
$2N-2n-2$ monopoles become mutually local and massless,
the $SO(2N)$ Seiberg-Witten curve has double zeros \cite{EOT}\cite{9712005} as,
\begin{equation}
\left[P_{2N}\left(x^2,u_i\right)\right]^2-4x^4\Lambda^{4N+4}
=x^2\left[H_{2N-2n-2}\left(x^2\right)\right]^2 F_{4n+2}(x).
\label{SO local monopole}
\end{equation}
For $Sp(2N)$ gauge group, 
$2N-2n$ monopoles become massless and Seiberg-Witten curve has 
double zeros \cite{9712149} as,
\begin{equation}
\left[P_{2N}\left(x^2,u_i\right)+2\Lambda^{2N+2}\right]^2-4\Lambda^{4N+4}
=\left[H_{2N-2n}\left(x^2\right)\right]^2 F_{4n}(x).
\label{Sp local monopole}
\end{equation}
Thus the exact superpotential in the confining phase is evaluated from  
the Seiberg-Witten curve with the massless monopole constraints 
(\ref{SO local monopole})(\ref{Sp local monopole}).

%%%%%%%%%%%%%%%%%%%%%%%%%%%%%%%%%%%%%%%%%%%%%%%%%%%%%%%%%%%%%%%%%%%%%%%%
\subsection{Confining Phase Superpotential at Large $N$}

We have seen that the exact confining phase superpotential 
is evaluated from the Seiberg-Witten theory.
Next we will consider its large $N$ limit.
In order to examine the duality, 
we need to evaluate 
the exact confining phase superpotential 
for $SO(N)$ and $Sp(N)$ gauge group in the large $N$ limit.
In this subsection, we will show that a solution for 
the massless monopole constraints 
(\ref{SO local monopole})(\ref{Sp local monopole})
of the $SO(2KN-2K+2)/Sp(2KN+2K-2)$ gauge group is found from 
that of $SO(N)/Sp(N)$ gauge group via Chebyshev polynomials
\cite{DS}.
%The solution of $P_{KN}(x)$ for the massless monopoles constraint
%to the $SO(2KN-2K+2)/Sp(2KN+2K-2)$ gauge group is obtained as follows.

%For the gauge group $SO(2N)$ with the classical superpotential 
%$W_{{\rm cl}}(\Phi)$, in the classical vacuum 
%$\Phi=\sigma_2 \otimes 
%\mathrm{diag}(ia_1\mathbf{1}_{N_0},\cdots,ia_r\mathbf{1}_{N_r})$ 
%superpotential is given by
%\begin{eqnarray}
%W_{{\rm cl}}=-\frac{N_1 m^2}{2 g} . \label{classical potential}
%\end{eqnarray}

%For the gauge group $SO(2N)$ with the classical superpotential 
%$W_{{\rm cl}}(\Phi)$, the gauge symmetry breaks as
%\begin{equation}
%SO(2N)\to SO(2N_0)\otimes_{i=1}^{n}U(N_i)
%\end{equation}
%where $N_i$'s satisfy $N_0+\sum_{i=1}^{n}N_i=N$.
For the gauge group $SO(2KN-2K+2)$ with the classical superpotential 
$W_{{\rm cl}}(\Phi)$, the gauge group breaks in the classical vacuum as,
\begin{equation}
SO(2KN-2K+2)\to SO(2KN_0-2K+2)\otimes_{i=1}^{n}U(KN_i).
\label{SO(KN) break} 
\end{equation}
where $N_i$'s satisfy $N_0+\sum_{i=1}^{n}N_i=N$.
We choose $P_{2KN-2K+2}$ such as,
\begin{equation}
P_{2KN-2K+2}(x)=\widetilde{\Lambda}^{4KN-4K}x^2
T_{K}\left(\frac{P_{2N}(x)}{x^2\Lambda^{2N-2}}\right),
\label{chev so}
\end{equation}
where  Chebyshev polynomials $T_K(x)$ and $U_{K}(x)\ (K=0,1,2,\cdots)$ 
are defined as,
\begin{eqnarray}
&&
T_K(x)\equiv\cos(K\arccos x),\quad 
U_{K}(x)\equiv \sin(K\arccos x),
\\
&&
T_K(x)^2-4=(x^2-4)U_{K-1}(x)^2.
\end{eqnarray}
Then this $P_{2KN-2K+2}$ satisfies the massless monopole condition 
for $SO(2KN-2K+2)$ gauge theory as,
\begin{eqnarray}
&&
P_{2KN-2K+2}(x)^2-4x^4\widetilde{\Lambda}^{4KN-4K}
\nn \\
&=&\widetilde{\Lambda}^{4KN-4K}\Lambda^{-4N+4}
\Bigg[U_{K-1}\left(\frac{P_{2N}(x)}{x^2\Lambda^{2N}}\right)\Bigg]^2
\left(P_{2N}(x)^2-4x^4\Lambda^{4N-4}\right)
\nn \\
&\equiv&x^2\left[H_{2KN-2n-2}(x)\right]^2F_{4n+2}(x).
\end{eqnarray}
Thus we found a solution of the massless 
monopole constraint for $SO(2KN-2K+2)$ gauge theory.

%For the gauge group $Sp(2N)$, 
%the exact superpotential can be analyzed in the same manner.
%%In the vacuum $\Phi=\sigma_3 \otimes 
%\mathrm{diag}(ia_1\mathbf{1}_{N_0},\cdots,ia_r\mathbf{1}_{N_r})$, 
%the classical superpotential is given by
%\begin{equation}
%W_{{\rm cl}}=-\frac{N_1 m^2}{2g},
%\end{equation}

For the gauge group $Sp(2N)$, 
the exact superpotential can be analyzed in the same manner. 
%In the classical vacua, the gauge group breaks as,
%\begin{equation}
%Sp(2N)\to Sp(2N_0)\otimes_{i=1}^{n}U(N_i)
%\end{equation}
%where $N_i$'s satisfy $N_0+\sum_{i=1}^{n}N_i=N$.
In the classical vacuum, 
the gauge group $Sp(2KN+2K-2)$ breaks in the classical vacuum as,
\begin{equation}
Sp(2KN+2K-2)\to Sp(2KN_0+2K-2)\otimes_{i=1}^{n}U(KN_i).
\label{Sp(KN) break} 
\end{equation}
If we choose $P_{2KN+2K-2}(x)$ as,
\begin{equation}
P_{2KN+2K-2}(x)=\frac{\widetilde{\Lambda}^{2KN+2K}}{x^2}
T_{K}\left(\frac{x^2P_{2N}(x)}{\Lambda^{2N+2}}+2\right)
-2\frac{\widetilde{\Lambda}^{2KN+2K}}{x^2},
\label{chev sp}
\end{equation}
this satisfies the massless
monopole constraint for the $Sp(2KN+2K-2)$ gauge theory as follows.
\begin{eqnarray}
&&
\left(x^2P_{2KN+2K-2}(x)+2\widetilde{\Lambda}^{2KN+2K}\right)^2
-4\widetilde{\Lambda}^{4KN+4K}
\nn \\
&=&
\widetilde{\Lambda}^{4KN+4K}\Lambda^{-4N-4}
\Bigg[
U_{K-1}\left(\frac{P_{2N}(x)^2}{\Lambda^{2N+2}}
\right)\Bigg]^2
\left(
\left(x^2P_{2N}(x)+2\Lambda^{2N+2}\right)^2-4\Lambda^{4N+4}
\right)
\nn \\
&\equiv&
\left[H_{2KN-2n}(x)\right]^2F_{4n}.
\end{eqnarray}
Thus a solution of the massless 
monopole constraint for $Sp(2KN+2K-2)$ gauge theory 
is also expressed in terms of the Chebyshev polynomial.

To evaluate the exact superpotential, 
we need to find the $\langle \tilde{u}_k \rangle$ 
for $SO(2KN-2K+2)/Sp(2KN+2K-2)$ gauge theory.
By expanding out (\ref{chev so}), the vacuum expectation values 
$\langle \tilde{u}_k \rangle$ for $SO(2KN-2K+2)/Sp(2KN+2K-2)$ gauge theory 
are related with the vacuum expectation values $\langle u_k \rangle$ 
for $SO(2N)/Sp(2N)$ as,
\begin{eqnarray}
&&\tilde{u_2}=Ku_2,\quad
\tilde{u_4}=Ku_4,\nn \\
&&
\tilde{u_6}=Ku_6+(K^2-K^3)\frac{u_2^3}{6}, \cdots.
\end{eqnarray}
When we consider the quartic classical superpotential 
$W_{{\rm cl}}(x)=gx^4/4+mx^2/2$, 
the exact superpotential for 
$SO(2KN-2K+2)/Sp(2KN+2K-2)$ gauge theory 
can be expressed in terms of that of $SO(N)/Sp(N)$ as,
\begin{equation}
W_{{\rm exact}}(\tilde{u}_i,g_i)=KW_{{\rm exact}}(u_i,g_i).
\end{equation}

For the completeness, we will consider the classical quartic superpotentials
for $SO(2KN-2K+2)/Sp(2KN+2K-2)$ gauge theory.
%in order to check that $SO(2KN-2K+2)/Sp(2KN+2K-2)$ gauge theory is
%really discussed from the $SO(2N)/Sp(2N)$ gauge theory.
%in order to check that $SO(2KN-2K+2)/Sp(2KN+2K-2)$ gauge theory is
%really discussed from the $SO(2N)/Sp(2N)$ gauge theory,
%we will consider the classical superpotential in the vacuum. 
For the gauge group $SO(2N)$ the classical quartic superpotential  
$W_{{\rm cl}}(\Phi)$ is evaluated in the classical vacuum as,
\begin{eqnarray}
W_{{\rm cl}}^{SO(2N)}=-\frac{N_1 m^2}{2 g}. \label{classical potential}
\end{eqnarray}
For the gauge group $Sp(2N)$, 
the classical quartic superpotential $W_{{\rm cl}}$ is given 
in the vacuum as,
\begin{equation}
W_{{\rm cl}}^{Sp(2N)}=-\frac{N_1 m^2}{2g}.
\label{classical sp}
\end{equation}
Since the gauge groups break as (\ref{SO(KN) break})(\ref{Sp(KN) break})
in the vacuum,
the classical quartic superpotential $\widetilde{W}_{{\rm cl}}$ for 
the $SO(2KN-2K+2)/Sp(2KN+2K-2)$ gauge theory is also written as 
$\tilde{W}_{{\rm cl}}=KW_{{\rm cl}}$.

%Thus the classical superpotential 
%$\widetilde{W}_{{\rm cl}}$ in the 
%vacuum for the $SO(2KN-2K+2)/Sp(2KN+2K-2)$ gauge theory
%is written as $\tilde{W}_{{\rm cl}}=KW_{{\rm cl}}$.

On the other hand, the factorization property of 
the effective superpotential in the deformed geometry 
is considered as follows.
Before the geometric transition, 
$SO(2KN-2K+2)$ gauge theory is realized by wrapping 
$2NK-2K+2$ D$5$-branes on $\mathbf{RP^2}$\footnote{
In this paper, we call $O^{-}$-plane (resp. $O^{+}$-plane)
as orientifold plane with negative (resp. positive) R-R charge.
}
around the exceptional 2-cycles 
in the resolved geometry.
Therefore the effective superpotential $W_{{\rm eff}}(S_i)$ 
generated by the $3$-form flux in the dual theory is  
evaluated as,
\begin{eqnarray}
&&W_{{\rm eff}}^{SO(2NK-2K+2)}=
\left((2N_0K-2K+2)-2\right)\Pi_0+\sum_{i=1}^n(2KN_i)\Pi_i
\nn \\
&=&KW_{{\rm eff}}^{SO(2N)}.
\end{eqnarray}
In the same way, 
the $Sp(2KN+2K-2)$ gauge theory is realized by wrapping 
$2NK+2K-2$ D$5$-branes and a $O^{+}$-plane around the exceptional 2-cycles 
in the resolved geometry.
The effective superpotential $W_{{\rm eff}}(S_i)$ in the dual 
theory is evaluated as,
\begin{eqnarray}
&&W_{{\rm eff}}^{Sp(2NK+2K-2)}=
\left((2N_0K+2K-2)+2\right)\Pi_0+\sum_{i=1}^n(2KN_i)\Pi_i
\nn \\
&=&KW_{{\rm eff}}^{Sp(2N)}.
\end{eqnarray}
Thus the factorization property holds for the effective superpotential 
in the deformed geometry.

In this way, exact superpotential for $SO(2KN-2K+2)/Sp(2KN+2K-2)$ gauge theory 
with the quartic classical superpotential 
is expressed via that of $SO(2N)/Sp(2N)$ gauge theory.
Using this analysis, we can discuss the large $N$ exact superpotential 
by taking the limit $K\to\infty$ 
and the large $N$ duality can be examined by 
checking the coincidence of the superpotentials 
for the {\it finite} rank gauge group.
%In the rest of this  section, we will evaluate the exact superpotential 
%$W_{{\rm low}}$ with classical quartic superpotential explicitly.
%and then 
%compare with the effective superpotential $W_{{\rm eff}}$
%generated by the $3$-form flux in the dual deformed geometry.

%%%%%%%%%%%%%%%%%%%%%%%%%%%%%%%%%%%%%%%%%%%%%%%%%%%%%%%%
\subsection{Computation of Confining Phase Superpotentials}

In this subsection, we will evaluate the confining phase superpotentials
for finite rank gauge groups
in terms of the gauge theoretical analysis. 
%At first, we will consider the case of $SO(2N)$ gauge theory 
%and then consider the case of $Sp(2N)$ gauge theory.
Although the coincidence should be hold for any $N$,
we will concentrate on some non-trivial examples as 
$SO(6)$, $SO(8)$ and $Sp(4)$ gauge theories in this paper.

\vspace{0.5cm}
\noindent{\bf Case 1:} $\mathbf{SO(6)\to SO(4)\times U(1)}$
\vspace{0.5cm}

In this case characteristic polynomial $P_6(x)$ 
which satisfies the constraint (\ref{SO local monopole}) is given by
\begin{eqnarray}
P_6=x^4(x^2-b^2)-2\Lambda^4x^2.
\end{eqnarray}
Using the formula (\ref{Newton}), we obtain the following relations, 
\begin{eqnarray}
u_2=b^2\qquad u_4=\frac{b^4}{2}+2\Lambda^4.
\end{eqnarray}
Thus the low energy superpotential is obtained as,
\begin{eqnarray}
W_{\mathrm{exact}}=\frac{m^2}{g}\left[\frac{1}{2}\left(\frac{gb^2}{m}\right)^2+2t^2+\left(\frac{gb^2}{m}\right)  \right].
\end{eqnarray}
Integrating out $b$, we can get the exact superpotential,
\begin{eqnarray}
W_{\mathrm{exact}}=-\frac{m^2}{2g}+\frac{2m}{g}t^2.
\label{so(6) pot}
\end{eqnarray}
where $t\equiv g\Lambda^2/m$. 
The gauge symmetry breaking can be read off in the classical limit, 
$\Lambda \to 0$. 
Comparing the above result with (\ref{classical potential}), 
we find $N_1=1$.\footnote{
In this subsection, we consider the brane configuration as 
$2N_0$ D$5$-branes wrapping on $\mathbf{RP^2}$ and 
$N_1$ D$5$-branes wrapping on each ${\bf P}^1$'s.
Therefore we are considering the gauge symmetry breaking as
$SO(2N)\to SO(2N_0)\times U(N_1)$.
} 
Using the relation $2N=2N_0+2N_1$ and $2N=6$, 
we obtain $2N_0=4$. 
Thus we found the exact superpotentials 
corresponding to the breaking as $SO(6)\to SO(4)\times U(1)$.

\vspace{0.5cm}
\noindent{\bf Case 2:} {\bf Splitting of ${\bf SO(8)}$}
\vspace{0.5cm}

Similarly we will analyze the gauge group $SO(8)$. 
In this case, we need to solve (\ref{SO local monopole}) for $2N=8$, 
%i.e. to find $P_8(x)$ such that
\begin{eqnarray}
P_8^2(x)-4\Lambda^{12}x^4=x^2 \left[H_4(x) \right]^2F_4(x). 
\label{so8 massless monopole}
\end{eqnarray}
Let us set\footnote{
Here we choose this particular ansatz for $H_4$ 
in order to avoid considering
the gauge symmetry breaking as $SO(8)\to SO(2)\times U(3)$.
In the later discussion, the expansion parameter $T$ for the 
effective superpotential of the deformed geometry is found 
to be ill-defined for this case.
}
$H_4(x)=x^2(x^2-a^2)$
and $P_8(x)=x^8+s_2x^6+s_4x^4+s_6x^2+s_8$. 
The condition (\ref{so8 massless monopole}) gives us following relations,
\begin{eqnarray}
s_8=0,\qquad s_6^2=4\Lambda^{12}, \qquad s_4=-3a^4-2s_2a^2, \label{s4} \\
a^4(2a^2+s_2)=\pm 4\Lambda^6.\qquad \qquad \label{so8 constraint}
\end{eqnarray}
Here we introduce new variable $b^2\equiv 2a^2+s_2$. 
Using this, we can rewrite (\ref{s4}) and (\ref{so8 constraint}) as,
\begin{eqnarray}
s_4=a^4-2a^2b^2, \qquad a^4b^2=\pm 4\Lambda^6. \label{constraint}
\end{eqnarray}
By the Newton's relation (\ref{Newton}), the Casimirs are now found as,
\begin{eqnarray}
u_2=2a^2-b^2, \qquad u_4=a^4+\frac{1}{2}b^4,
\end{eqnarray}
and the low energy superpotential finally takes the form,
\begin{eqnarray}
W_{\mathrm{exact}}&=&gu_4+mu_2+\beta(a^4b^2 \pm 4\Lambda^6) \nn\\
{}&=&\frac{m^2}{g}\left[x^4+\frac{1}{2}y^4+2x^2-y^2+\gamma (x^4y^2\pm 4t^3) 
\right],
\label{exact pot SO(8)1}
\end{eqnarray}
where $\beta$ and $\gamma \equiv \beta m/g^2$ are Lagrange multipliers, 
and  $x,y$ are dimensionless variables defined by $a^2=mx^2/g, b^2=my^2/g$. 
To get the low-energy superpotential, we want to integrate out $x,y$. 
Therefore we have to solve 
$\partial W_{\mathrm{exact}}/\partial x=0, 
 \partial W_{\mathrm{exact}}/\partial y=0$. 
Eliminating the Lagrange multipliers, 
we obtain the following equation and constraint for $x,y$.
\begin{eqnarray}
x^4+x^2-y^4+y^2=0, \nn \\
x^4y^2=\pm 4t^3. \qquad \label{so8}
\end{eqnarray}
From the above relations,
we can see how two different splittings will come out. 
In the classical limit $\Lambda\to 0$, 
the relations can be solved in two ways, namely, $x=0$ or $y=0$. 
Plugging these solutions into the superpotential, 
the former case corresponds to the gauge symmetry breaking 
$SO(6)\times U(1)$ and the latter case corresponds to
that of $SO(4)\times U(2)$.
%\newpage

\vspace{0.5cm}
$\mathbf{SO(8)\to SO(6)\times U(1)}$
\vspace{0.5cm}

First, we will consider the solution which become $x=0$ in the classical 
limit.
The equations (\ref{so8}) are rewritten as, 
\begin{eqnarray}
\frac{4t^3}{y^2}+\frac{2t^{\frac{3}{2}}}{y}-y^4+y^2=0.
\end{eqnarray}
This equation can be solved recursively using $t^{\frac{3}{2}}$ 
as expansion parameter. 
Plugging this solution into (\ref{exact pot SO(8)1}), 
we obtain the low energy superpotential for this case,
%Once the solution is obtained, $y$ can also be found and plugging them back 
%into the superpotential, we have 
\begin{eqnarray}
W_{\mathrm{exact}}=\frac{m^2}{g}\left[-\frac{1}{2}+4t^{\frac{3}{2}}+2(t^{\frac{3}{2}})^2-2(t^{\frac{3}{2}})^3+4(t^{\frac{3}{2}})^4-21(t^{\frac{3}{2}})^5+\mathcal{O}\left((t^{\frac{3}{2}})^6  \right)     \right]. \label{lowpote}
\end{eqnarray}
As $SO(6)$ case, we can read off  the gauge symmetry breaking pattern from 
the classical limit of this potential as $SO(8)\to SO(6)\times U(1)$.

\vspace{0.5cm}
$\mathbf{SO(8)\to SO(4)\times U(2)}$
\vspace{0.5cm}

Next, we will consider the solution which become $y=0$ in the classical 
limit.
In this case,
the equations (\ref{so8}) are rewritten as,
\begin{eqnarray}
x^4+x^2-\frac{16t^6}{x^8}-\frac{t^3}{x^4}=0 .
\end{eqnarray}
We can solve this equation as before but using as expansion parameter $t^3$. 
Plugging this back in $W_{{\rm exact}}$, we get the following superpotential.
\begin{eqnarray}
W_{\mathrm{exact}}=\frac{m^2}{g}\left[-1+4t^3-8t^6+64t^9-768t^{12}+\mathcal{O}(t^{13}) \right]. \label{low3}
\end{eqnarray}
In the classical limit, the gauge symmetry breaking pattern is found as,
$SO(8)\to SO(4)\times U(2)$.

%These two results (\ref{lowpote}), (\ref{low3}) are agreement 
%with the dual geometry result to this order .

%So that as a example we consider the case $Sp(4) \to Sp(2)\times U(1)$.

\vspace{0.5cm}
\noindent{\bf Case 3:} $\mathbf{Sp(4)\to Sp(2)\times U(1)}$
\vspace{0.5cm}

As an example of $Sp(2N)$ gauge theory, 
we will consider $Sp(4)$ gauge group.
The massless monopole condition (\ref{Sp local monopole})
becomes in this case as,
\begin{eqnarray}
(x^2P_{4}(x)+2t^3)^2-4t^6=(x^2-a^2)^2F_8(x).\label{sp mono}
\end{eqnarray}
Let us set $P_4(x)=x^4+s_2x^2+s_4$. 
The above condition gives us following equations .
\begin{eqnarray}
s_4=-3a^4-2s_2a^2, \qquad a^4(2a^2+s_2)=4t^3. \label{sp eq}
\end{eqnarray}
Introducing new variable $b^2\equiv 2a^2+s_2$, 
the equation (\ref{sp eq}) is rewritten as,
\begin{eqnarray}
s_2=b^2-2a^2,\qquad s_4=a^4-2a^2b^2, \qquad a^4b^2=4t^3.
\end{eqnarray}
Using the Newton's relation (\ref{Newton}), we have the relations as,
\begin{eqnarray}
u_2=2a^2-b^2,\qquad u_4=a^4+\frac{b^4}{2}.
\end{eqnarray}
Under the constraint (\ref{sp mono}), 
the low energy superpotential is written as,
\begin{eqnarray}
W_{\mathrm{exact}}=\frac{m^2}{g}\left[x^4+\frac{y^4}{2}+2x^2-y^2+\beta \left(x^4 y^2-4t^3 \right)\right].
\end{eqnarray}
where $\beta$ is a Lagrange multiplier and $x,y$ are dimensionless 
variable defined by $a^2=mx^2/g,b^2=my^2/g$. 
As $SO(2N)$ case, we have to integrate out $x,y$. 
After some calculations, we obtain the following equation.\footnote{
Here we remark that this equation is same as that of 
$SO(8) \to SO(6)\times U(1)$.
This is consistent with the analysis in the dual geometry.
The effective superpotential (\ref{superfulx})
of $N_0=1$, $N_1=1$ for negative orientifold plane charge is equal to that of
$N_0=3$, $N_1=1$ for positive orientifold plane charge.
Therefore these theories should have same low-energy superpotentials 
$W_{{\rm exact}}$. }
\begin{eqnarray}
\frac{4t^3}{y^2}+\frac{2t^{\frac{3}{2}}}{y}-y^4+y^2=0.
\end{eqnarray}
Therefore, the low energy superpotential is give by
\begin{eqnarray}
W_{\mathrm{exact}}=\frac{m^2}{g}\left[-\frac{1}{2}+4t^{\frac{3}{2}}+2(t^{\frac{3}{2}})^2-2(t^{\frac{3}{2}})^3+4(t^{\frac{3}{2}})^4-21(t^{\frac{3}{2}})^5+\mathcal{O}\left((t^{\frac{3}{2}})^6  \right)     \right] \ . \label{low2}
\end{eqnarray}
%As in the case of $SO(2N)$, we can read the breaking pattern from the
%classical limit of this result. 
By comparing this result with (\ref{classical sp}) in the classical limit,
we find $N_1=1$ and $N_0=1$. 
Thus this low-energy superpotential corresponds to the 
gauge symmetry breaking $Sp(4)\to Sp(2)\times U(1)$.

%%%%%%%%%%%%%%%%%%%%%%%%%%%%%%%%%%%%%%%%%%%%%%%%%%%%%%%%%%%%
\section{Analysis of Dual Geometry}
%%%%%%%%%%%%%%%%%%%%%%%%%%%%%%%%%%%%%%%%%%%%%%%%%%%%%%%%%%%%
%As discussed in the analysis of the gauge theory,
%we will concentrate on the case of $n=1$ and  
%Eq (\ref{classical sup}) becomes as,
%\begin{eqnarray}
%W(x)&=& gu_4+mu_2, \\
%W^{\prime}(x)&=&g\ x \left(x^2+a_1^2 \right),\qquad a_1=\sqrt{\frac{g}{m}}.
%\end{eqnarray}

As reviewed in section 2, the 3-form fluxes through the special Lagrangian 
$3$-cycles generate the superpotential and it can be identified with
the effective superpotential in the dual gauge theory.
In this section, we will calculate the periods $S_i$'s and $\Pi_i$'s for the 
case of $n=1$.
% and evaluate the effective superpotential $W_{{\rm eff}}$.
To find effective superpotential for the glueball superfields, 
we will compute the dual periods $\Pi_i$'s in terms of the periods 
$S_i$'s.

%%%%%%%%%%%%%%%%%%%%%%%%%%%%%%%%%%%%%%%%%%%%%%%%%%%%%%%%%%
\subsection{Monodoromy Analysis}

As in \cite{CIV}, we will express the period of the dual cycles 
in terms of the period of $S^3$'s. 
In this subsection, we will discuss their logarithmic terms 
from monodromy analysis.
First, we consider the transformation, 
$\Lambda_0 \to e^{2\pi i}\Lambda_0$. 
Under this transformation, the period $\Pi_i$ changes by,
\begin{eqnarray}
\Delta \Pi_i=-2 \left(S_0+2S_1 \right).
\end{eqnarray}
Therefore, we have the logarithmic dependence on $\Lambda_0$ as,
\begin{eqnarray}
\Pi_i=-\frac{2}{2\pi i}\left(S_0+2S_1 \right) \log \Lambda_0+\cdots .
\end{eqnarray}

Next we consider the transformation, 
$\mu_i \to e^{2\pi i}\mu_i$ ($\mu_i \equiv S_i/2W^{\prime\prime}(a_i)$). 
Under this transformation, $\Pi_i$ changes by
\begin{eqnarray}
\Delta \Pi_i=S_i ,
\end{eqnarray}
so that,
\begin{eqnarray}
\Pi_i=\frac{S_i}{2\pi i}\log \frac{S_i}{2W^{\prime\prime}(a_i)}.
\end{eqnarray}
 
%%%%%%%%
 \begin{figure}[ht]
 \centerline{\epsfxsize=7cm \epsfbox{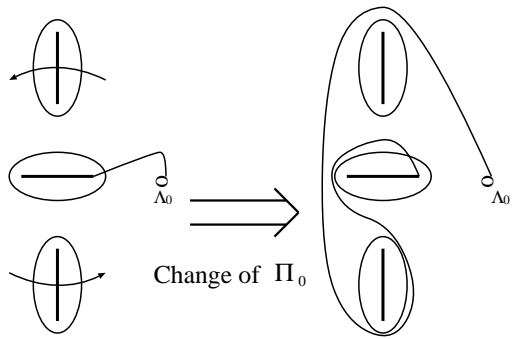}}
  \vskip 3mm
  \caption{The change of $\Pi_0$ under the $a_1\to e^{2\pi i}a_1$}
  \label{monod}
 \end{figure}
%%%%%%%%

%\begin{figure}[htbp]
%\hspace{4cm}
% \centerline{\epsfxsize=6cm \epsfbox{IIBmonodromy.eps}}
%	
%	\caption{The change of $\Pi_0$ under the $a_1\to e^{2\pi i}a_1$}
%	\label{monod}
%\end{figure}

Finally, we will consider the transformation, 
$a_1 \to e^{2\pi i}a_1$(see Fig\ref{monod}). 
Under this transformation $\Pi_0,\Pi_1$ change respectively by
 \begin{eqnarray}
 \Delta \Pi_0=2\left(2S_1 \right), \qquad \Delta \Pi_1=2\left(S_0+S_1 \right).
 \end{eqnarray}

Collecting these results, we have the following monodromy contributions.
\begin{eqnarray}
\Pi_0&=&\frac{W(\Lambda_0)-W(0)}{\pi i}
-\frac{1}{\pi i}\left(S_0+2S_1 \right)\log \Lambda_0
\nn \\
&&
+\frac{S_0}{2\pi i}\log \frac{S_0}{2W^{\prime\prime}(0)}
+\frac{4}{2\pi i}S_1 \log a_1+\cdots, 
\nn \\
\Pi_1&=&\frac{W(i\Lambda_0)-W(ia_1)}{\pi i}
-\frac{1}{\pi i}\left(S_0+2S_1 \right)\log \Lambda_0
\nn \\
&&
+\frac{S_1}{2\pi i}\log \frac{S_1}{2W^{\prime\prime}(ia_1)}
+\frac{2}{2\pi i}\left(S_0+S_1\right) \log a_1+\cdots. \label{monodterm}
\end{eqnarray}
Plugging these terms into (\ref{superfulx}), 
we obtain the effective superpotential, which is derived 
from naive integrating in \cite{Yank}\cite{CIV}.

%%%%%%%%%%%%%%%%%%%%%%%%%%%%%%%%%%%%%%%%%%%%
\subsection{Effective Superpotential}

In the previous section, 
we obtained the logarismic contribution of the dual periods $\Pi_i$'s
from the monodromy analysis.
In this subsection we will compute the remaining terms. 
The explicit computation of $\Pi_0$ and $\Pi_1$ can be found in Appendix A 
up to the 4th order in $S_i$. 
%Here we will only show the result that will be used later. 
\begin{eqnarray}
-\pi i \Pi_0&=&-W(\Lambda_0)+W(0)-
\left(S_0+2S_1 \right)\log \Lambda_0
\nn \\
&&
+\frac{S_0}{2}\log \frac{S_0}{2W^{\prime\prime}(0)}
+2S_1 \log a_1
\nn \\
&&
+ga_1^4 \Bigg[-2 \left(x+y \right)\log \left(\frac{\Lambda_0}{a_1} \right)
+x(\log x -1)
\nn \\
&&
+4xy-\frac{3}{2}x^2-\frac{1}{2}y^2
+\frac{9}{2}x^3 
-21 x^2y+9 x y^2 -\frac{1}{2} y^3 
\nn \\
&&
- \frac{45}{2} x^4
+\frac{466}{3} x^3 y+\frac{76}{3} x y^3-131x^2y^2-\frac{5}{6}y^4  \Bigg]
+\cdots,
\label{Pi0}\\
-\pi i \Pi_1&=&-W(i\Lambda_0)+W(ia_1)
-\left(S_0+2S_1 \right)\log \Lambda_0
\nn \\
&&
+\frac{S_1}{2}\log \frac{S_1}{2W^{\prime\prime}(ia_1)}
+2\left(S_0+S_1\right) \log a_1
\nn \\
&&
+ga_1^4 
\Bigg[ -2(x+y)\log \left(\frac{\Lambda_0}{a_1} \right)
+\frac{y}{2}\left(\log y-1 \right)
\nn \\
&&
+2x^2-xy-7x^3+9x^2y-\frac{6}{4}xy^2
\nn \\
&&
+\frac{233}{6}x^4
-\frac{262}{3}x^3y+\frac{152}{4}x^2y^2-\frac{10}{3}xy^3 \Bigg]+\cdots,
\label{Pi1}
\end{eqnarray}
where $x \equiv S_0/2ga_1^4$ , $y \equiv S_1/ga_1^4$.  
In this expression, the cut off $\Lambda_0$ is combined 
with the bare coupling $\alpha$ to form
the gauge theory scale $\Lambda$ 
of the underlying $\mathcal{N}=2$ Yang-Mills theory \cite{CIV}\cite{EOT}.

The effective superpotential is given by (\ref{superfulx}). 
%Since we want to compare this with the field theory result, 
%we need to integrate out $x,y$ using 
Since the exact low energy superpotential is obtained by integrating out 
the glueball superfields $S_0,S_1$,
we need to solve the equations
$\partial W_{{\rm eff}}/\partial x=0\ , 
\ \partial W_{{\rm eff}}/\partial y=0$. 
In the leading order, these equations are solved as,
\begin{eqnarray}
x=T^{\frac{N_1}{2}},\qquad y=T^{\frac{N_0}{2}-1}, \qquad 
T\equiv \left(\frac{\Lambda}{a_1} \right)^{\frac{4(N_0+2N_1-2)}{N_1(N_0-2)}}.
\end{eqnarray}
Using $T$ as the expansion parameter, 
the low energy superpotential $W_{\mathrm{low}}$ can be evaluated. 
Now we will compute some examples that correspond to the examples
in the previous section.

\vspace{0.5cm}
$\mathbf{SO(6)\to SO(4)\times U(1)}$
\vspace{0.5cm}

As the first example, we consider the case $SO(6)\to SO(4)\times U(1)$. 
This gauge theory is realized by taking values as, $N_0=4\ , \ N_1=1$ 
and $T=\frac{g^2\Lambda^4}{m^2}\equiv t^2$
The effective superpotential is obtained by plugging these values 
into the expression (\ref{superfulx}). 
%Thus the effective superpotential is given by,
\begin{eqnarray}
W_{\mathrm{low}}=\frac{m^2}{g}\left[-\frac{1}{2}+2T+\mathcal{O}(T^5) \right].
\end{eqnarray}
In the calculation of this low-energy superpotential, 
some miraculous cancellation happens and 
this superpotential coincides with the confining phase superpotential 
(\ref{so(6) pot}).
Thus the large $N$ duality is proved for this case up to the order
${\cal O}(T^5)$ .

\vspace{0.5cm}
\noindent{\bf Case 2: Splitting of ${\bf SO(8)}$}
\vspace{0.5cm}

As a next example, we will consider the case which corresponds to the 
splitting of the gauge group $SO(8)$.
In this case, we consider the superpotentials generated by 
the fluxes of $N=N_0+2N_1=8$ D5-branes and a $O^{-}$-plane.
To compare with the results of gauge theory, we will consider the 
following two breaking patterns.

\vspace{0.5cm}
$\mathbf{SO(8)\to SO(6)\times U(1)}$
\vspace{0.5cm}

First, we consider the case $SO(8)\to SO(6)\times U(1)$.
This breaking is realized by choosing 
$N_0=6\ , \ N_1=1$ and $T=t^{\frac{3}{2}}$. 
Thus effective superpotential is given by,
\begin{eqnarray}
W_{\mathrm{low}}=\frac{m^2}{g}\left[-\frac{1}{2}+4T+2T^2-2T^3+4T^4+\mathcal{O}(T^5) \right].
\end{eqnarray}
This superpotential coincides with 
the exact superpotential (\ref{lowpote}) in the gauge theory analysis 
up to ${\cal O}(T^5)$.

\vspace{0.5cm}
$\mathbf{SO(8)\to SO(4)\times U(2)}$
\vspace{0.5cm}

As another splitting, we consider the case $SO(8)\to SO(4)\times U(2)$.
This breaking corresponds to the case 
$N_0=4\ , \ N_1=2$ and $T=t^{\frac{3}{2}}$. 
Thus effective superpotential is 
\begin{eqnarray}
W_{\mathrm{low}}=\frac{m^2}{g}\left[-1+4T^2-8T^4+64T^6-768T^8+\mathcal{O}(T^{10}) \right].
\end{eqnarray}
This superpotential also coincides with the confining phase superpotential 
(\ref{low3}).
Thus we examined the large $N$ duality for both of the breaking patterns
in the gauge theory.

\vspace{0.5cm}
\noindent{\bf Case 3:} $\mathbf{Sp(4)\to Sp(2)\times U(1)}$
\vspace{0.5cm}

Finally we consider a example of $Sp$ gauge theory.
In this case, we have to choose the plus sign in (\ref{flux pot}) and 
set $N_0=2,N_1=1$. 
The low-energy superpotential is 
\begin{eqnarray}
W_{\mathrm{low}}=-\frac{m^2}{g}\left[-\frac{1}{2}+4T+2T^2-2T^3+4T^4+\mathcal{O}(T^5) \right].
\end{eqnarray}
%If we put $N_0+4$ in (\ref{flux pot}), 
%we can see the case of plus sign and $N_0$.
%Therefore this result is the same as $SO(8)\to SO(6)\times U(1)$. 
This superpotential coincides with the confining phase superpotential 
(\ref{low2})
Thus we examined the large $N$ duality for $Sp(2N)$ gauge theory.

%\section{Conclusion}
%In the present work, we have confirmed the large $N$ duality proposed 
%in \cite{CIV}\cite{EOT} for $SO/Sp$ gauge theories 
%by using the confining phase superpotential. 
%Although this duality make sense for large rank of gauge groups, 
%we can reduce the rank of gauge group to finite one in terms of 
%Chebyshev polynomials. 
%Thus we can check the duality explicitly 
%by using confining phase superpotential for finite rank gauge groups. 
%As an example we have considered 4 breaking patterns. 
%In these examples dual geometry analysis perfectly agrees 
%with the field theory analysis up to $\mathcal{O}(S_i^4)$. 

%\newpage
\vspace{1cm}
\begin{center}
{\large{\bf Acknowledgements}}
\end{center}

The authors thank to Norisuke Sakai for 
reading this manuscript and giving us useful comments.
The authors are obliged especially to Katsushi Ito for 
stimulus discussions and useful suggestions. 
Y.O. would like to thank M.Eto for continuous encouragement.
H.F. is supported by JSPS research fellowship for young scientists.

\setcounter{equation}{0}
\renewcommand{\theequation}{A.\arabic{equation}}
\appendix
\section*{Appendix A \, Computation of Periods }
%\chapter*{Appendix}
%\appendix
%\section*{Appendix A}{Computation of Periods }

In this appendix we will show the explicit computation of the periods 
$\Pi_0, \Pi_1$ in (\ref{superfulx}). 
%This computation is similar to $SU(N)$ case \cite{CIV}. 
As discussed in section 2, 
the effective one-form is written as follows,
\begin{equation}
2dx \sqrt{W'^2(x)+f_2(x)}=2dx g \sqrt{(x^2-{0^+}^2)(x^2+x_1^2)(x^2+x_2^2)}.
\end{equation}
Comparing the coefficient of $x^3$ on both sides in the above equation,
we obtain the following relation.
\begin{equation}
\frac{m}{g}=\frac{1}{2}\left(-\Delta_0^2+x_1^2+x_2^2 \right)=a_1^2.
\end{equation}
Here we define new variables given by,
\begin{eqnarray}
\Delta_0\equiv 0^+= -0^-,  \qquad \Delta_1 \equiv \frac{1}{2}(x_2-x_1), \\ 
Q\equiv \frac{1}{2}(x_2+x_1),\quad w\equiv \frac{\Delta_0}{Q},\ \ u\equiv \frac{\Delta_1}{Q}.
\end{eqnarray}
Since $f_2$ is considered as a small perturbation, 
$\Delta_0$, $\Delta_1$ and $Q$ satisfy
\begin{equation}
|\Delta_0 | \sim |\Delta_1 |\ll |Q|.
\end{equation}
%We will use this in order to expand all four periods in powers of $w$ and $u$.
Under this relation, we can expand the periods $S_0$ and $S_1$ in powers 
of $w$ and $u$.
\begin{eqnarray}
S_0 &=& \frac{g}{\pi i}\int_{-\Delta_0}^{\Delta_0} \sqrt{(x^2-\Delta_0^2)(x^2+x_1^2)(x^2+x_2^2)}\ dx \nn \\
{}&=&\frac{g}{2}Q^2 w^2 \left[1-u^2+\frac{w^2}{4}+\frac{w^2u^2}{2}+\frac{w^2u^4}{2}-\frac{u^2w^4}{4}+\mathcal{O}\left((uw)^7 \right) \right], \\ 
\label{period s0}
%\end{eqnarray}
%The computation for $S_1$ is obtained in a similar way.
%\begin{eqnarray}
S_1 &=& \frac{g}{ \pi i}\int_{x_1}^{x_2} \sqrt{(x^2-\Delta_0^2)(x^2+x_1^2)(x^2+x_2^2)}\ dx  \nn \\
{}&=&gQ^4u^2 \left[1+\frac{1}{2}w^2 -\frac{1}{8} w^4 - \frac{1}{8} w^4 u^2+\frac{1}{16} w^6+ \mathcal{O}\left((uw)^7 \right) \right]. \label{period s1}
\end{eqnarray}

%Let us compute the dual periods. 
%In this case we have to keep in mind that $\Lambda_0$ will be taken to infinity at the end and therefore we shall discard any contribution of order $\Lambda_0^{-1}$ or higher in an expansion around infinity. In this case it is useful to define a generating functions,
Next we will compute the dual periods $\Pi_i$'s.
In the computation, we will discard any contributions 
of ${\cal O}(\Lambda_0^{-1})$, since $\Lambda_0$ is introduced as
cut-off of infinite volume of dual three cycles.
%\begin{eqnarray}
%G(a)&\equiv& -\sqrt{\frac{x_1^2+\Delta_0^2+a}{x_1^2+a}} \log\left(\sqrt{x_1^2+\Delta_0^2+a}+\sqrt{x_1^2+a}-\log \Delta_0 \right),\nn \\
%H(a)&\equiv& -a\left(\log(4\Lambda_0^3)-\log (4Q^2u) \right) \nonumber \\
%&&+\sqrt{(x_1^2+a)(x_2^2+a)}
%\log \Bigg((\Lambda_0^3+a)(x_1^2-x_2^2)/
%\{-(x_1^2+x_2^2+2a)(\Lambda_0^3+a)\nn \\
%&&
%+2(x_1^2+a)(x_2^2+a)
%+2\sqrt{(x_1^2+a)(x_2^2+a)(\Lambda_0^3-x_1^2)(\Lambda_0^3-x_2^2)}\}\Bigg)
%\nn
%\end{eqnarray}
\begin{eqnarray}
\frac{\pi i}{g} \Pi_0&=&\int_{\Delta_0}^{\Lambda_0}\sqrt{(x^2-\Delta_0^2)(x^2+x_1^2)(x^2+x_2^2)}\ dx \nonumber \\ 
{}&=&\frac{1}{4}(\Lambda_0)^4+\frac{1}{2}(\Lambda_0)^2Q^2\left(1+u^2-\frac{1}{2}w^2 \right)\nn \\
&&
-Q^4 \log(2\Lambda_0)\left(\frac{w^2}{2}+\frac{w^2u^2}{2}+\frac{w^4}{8}+2u^2 \right) \nonumber \\
&&+\frac{Q^4}{32}w^4-\frac{Q^4w^2}{4}\left(1+u^2 \right)+\sum_{n=2}^{\infty}c_n \frac{(-1)^{n-2}}{(n-2)!}(4Q^2u)^nG^{(n-2)}(a)\Bigg|_{a=0}
 \label{period pi0}
\end{eqnarray}
where $c_n$ are the coefficient of the power expansion of $\sqrt{1+x}$. The computation for $\Pi_1$ is obtained in a similar way.
%\begin{eqnarray}
%\frac{\pi i}{g} \Pi_1&=&\int_{x_2}^{\Lambda_0}\sqrt{(x^2-\Delta_0^2)(x^2+x_1^2)(x^2+x_2^2)}\ dx \nonumber \\ 
%{}&=&g^{-1}W(i\Lambda_0)-2Q^2\left(u^2+\frac{w^2}{4}+\frac{w^2u^2}{4}+\frac{1}{16}w^4 \right)\log (\Lambda_0)\nn \\
%&&
%-\frac{Q^4}{4}\left(1+u^2 \right)+\frac{Q^4}{4}\left(1+u^4 \right) \nonumber \\
%&&+Q^4u^2 \left(1+\frac{w^2}{2}-\frac{w^4}{8}-\frac{w^4u^2}{8}+\frac{1}{16}w^6 \right)\log u \nn \\
%&&
%+Q^2\left(u^2+\frac{w^2}{4}+\frac{w^2u^2}{4}+\frac{1}{16}w^4 \right)\log (Q^2) \nonumber \\
%&&+\frac{Q^4}{64}w^6\left(1+3u^2 \right)+\frac{Q^4}{16}w^4-\frac{5Q^4}{1536}w^8 \label{period pi1}
%\end{eqnarray}
\begin{eqnarray}
\frac{\pi i}{g}\Pi_1&=&\int_{x_2}^{\Lambda_0}\sqrt{(x^2+\Delta_0^2)(x^2+x_1^2)(x^2+x_2^2)}dx  \nonumber \\
&=&\frac{\Lambda_0^6}{4}+\Lambda_0^3 Q^2 \left(-\frac{1+u^2}{2}+\frac{w^2}{4} \right)+Q^4 \log (4\Lambda^3) \left(-u^2-\frac{(1+u^2)w^2}{4} \right) \nonumber \\
{}&-&Q^4 \frac{(1+u^2)w^2}{4}+Q^4\left(u^2+\frac{w^2(1+u^2)}{4} \right)\log (4Q^2u^2)+Q^4\frac{1+u^4}{4} \nonumber \\
{}&-&\frac{1}{2}\sum_{n=1}^{\infty}c_n (-1)^{(n)}\frac{w^{2n}Q^{2n}}{(n-1)!}
H^{(n-1)}(a)\Bigg|_{a=0}.
\label{period pi1}
\end{eqnarray}
In the above expansion, we used the following generating functions 
\begin{eqnarray}
G(a)&\equiv& -\sqrt{\frac{x_1^2+\Delta_0^2+a}{x_1^2+a}} \log\left(\sqrt{x_1^2+\Delta_0^2+a}+\sqrt{x_1^2+a}-\log \Delta_0 \right),\nn \\
H(a)&\equiv& -a\left(\log(4\Lambda_0^3)-\log (4Q^2u) \right) \nonumber \\
&&+\sqrt{(x_1^2+a)(x_2^2+a)}
\log \Bigg((\Lambda_0^3+a)(x_1^2-x_2^2)/
\{-(x_1^2+x_2^2+2a)(\Lambda_0^3+a)\nn \\
&&
+2(x_1^2+a)(x_2^2+a)
+2\sqrt{(x_1^2+a)(x_2^2+a)(\Lambda_0^3-x_1^2)(\Lambda_0^3-x_2^2)}\}\Bigg).
\nn
\end{eqnarray}

%Using (\ref{period s0})$\sim$(\ref{period pi1}), 
%we can explicitly compare order by order in $w$ and $u$ 
%the two expressions for $\Pi_0$ and $\Pi_1$ given 
%by (\ref{period pi1}) and (\ref{monodterm}) to obtain the following result.
The dual periods 
$\Pi_0$ and $\Pi_1$ are expressed in terms of 
the periods $S_0$ and $S_1$ by comparing 
order by order in $w$ and $u$ in the expressions 
(\ref{period s0})$\sim$(\ref{period pi1}).
\begin{eqnarray}
-\pi i \Pi_0&=&
-W(\Lambda_0)+W(0)
- \left(S_0+2S_1 \right)\log \Lambda_0 +\frac{S_0}{2} \log \frac{S_0}{2W''(0)}+2S_1 \log a_1-\frac{S_0}{2} \nonumber \\
&&+\frac{1}{4ga_1^4}\left(8S_0 S_1-\frac{3}{2}S_0^2-2S_1^2 \right) +\frac{1}{8(ga_1^4)^2}\left(\frac{9}{2}S_0^3-42S_0^2S_1+36S_0S_1^2-4S_1^3 \right) \nonumber \\
&&+\frac{1}{16(ga_1^4)^3}\left(-\frac{45}{2}S_0^4+\frac{932}{3}S_0^3S_1+\frac{608}{3}S_0S_1^3-524S_0^2S_1^2-\frac{40}{3}S_1^4 \right)\nn \\
&&+\cdots, \\
-\pi i \Pi_1&=&
-W(i\Lambda_0)+W(ia_1)-\left(S_0+2S_1 \right)\log \frac{\Lambda_0}{a_1}-S_1 \log a_1 +\frac{S_1}{2} \log \frac{S_1}{m}-\frac{S_1}{2} \nonumber \\
&&+\frac{1}{4ga_1^4}\left(2S_0^2-2S_0S_1 \right)+\frac{1}{8(ga_1^4)^2}\left(-7S_0^3+18S_0^2S_1-6S_0S_1^2 \right) \nonumber \\
&&+\frac{1}{16(ga_1^4)^3}\left(\frac{233}{6}S_0^4-\frac{524}{3}S_0^3S_1+152S_0^2S_1^2-\frac{80}{3}S_0S_3^3 \right)+\cdots.
\end{eqnarray}

\end{document}